\documentclass{article}
\usepackage[utf8]{inputenc}
\usepackage{graphicx}
\usepackage[colorlinks=true, allcolors=blue]{hyperref}
\usepackage{booktabs}
\usepackage{multirow}
\usepackage{pdflscape}
\usepackage{afterpage}
\usepackage{adjustbox}

\usepackage{amsmath,amssymb,amsthm}
\usepackage[square,numbers]{natbib}
\title{On fitting the Lomax distribution: a comparison between minimum distance estimators and other estimation techniques}

\author{T. Nombebe, J.S. Allison, L. Santana, \& I.J.H. Visagie\footnote{E-mail:\texttt{jaco.visagie@nwu.ac.za}}\\
School of Mathematical and Statistical Sciences,\\ North-West University,\\ South Africa.\\
}

\date{\today}

\begin{document}

\maketitle
\begin{abstract}
In this paper we investigate the performance of a variety of estimation techniques for the scale and shape parameter of the Lomax distribution. These methods include traditional methods such as the maximum likelihood estimator and the method of moments estimator. A version of the maximum likelihood estimator adjusted for bias is also included. Furthermore, alternative moment-based estimation techniques such as the $L$-moment estimator and the probability weighted moments estimator are included along with three different minimum distance estimators. The finite sample performances of each of these estimators is compared via an extensive 
Monte Carlo study. We find that no single estimator outperforms its competitors uniformly. We recommend one of the minimum distance estimators for use with smaller samples, while a bias reduced version of maximum likelihood estimation is recommended for use with larger samples. 
{In addition, the desirable asymptotic properties of traditional maximum likelihood estimators make them appealing for larger samples.}
We also include a practical application demonstrating the use of the techniques on observed data.
\end{abstract}

\section{Introduction}

The Pareto distribution is a heavy-tailed distribution that was originally developed in the 19$^\textrm{th}$ century to model the distribution of income among individuals \citep{PARE1897}. However, in the years following its introduction, it has been extensively modified and changed to produce several variants, referred to as the Type I, II, III, and IV Pareto distributions as well as the so-called generalised Pareto distribution (GPD). The focus of this paper is on the Pareto Type II distribution where the location parameter is zero, also known as the \emph{Lomax} distribution. This rather popular distribution was originally introduced to model business failure data \citep{LOMA1954}, but has subsequently been employed as a model in a wide range of settings. 
{This distribution has also been found to be extremely useful in modelling survival times when censoring is present. In \cite{MITR2021} the lifetimes of electric power transformers are modelled and it is found that the Lomax distribution provides the best fit among the various distributions considered.}

{Another important application of the Lomax distribution is found in autoregressive conditional duration (ACD) models, which model high frequency financial data, where the duration between market events are of interest. Without discussing the details of the model we will briefly describe the role of the Lomax distribution in ACD models (for an in-depth exposition of ACD models, see \cite{ER1998} and \cite{Tha2010}). We only note that the error terms of the ACD model are historically assumed to be exponential or Weibull distributed, see, for example, \cite{MMO2020}. However, in \cite{BG2001} and \cite{BGGV2004}, it is found that these choices do not adequately describe the characteristics of the observed data. In \cite{DG2004}, the authors advocate the use of infinite mixtures of exponential distributions to model the distribution of the error term, finding that this choice improves the fit of the model.}

{If the scale parameter of this mixing exponential distribution follows an inverse Gaussian distribution, then the resulting distribution of the error term is Lomax. It is therefore important for the efficient implementation of these ACD models that we are able to accurately estimate the parameters of this distribution. Another important area of ongoing research is goodness-of-fit tests for the Lomax and related distributions as these tests check the assumptions underlying the ACD models.}

Other scenarios in which the Lomax distribution has been used as a model include wealth distribution \citep{ATHA1978},
the distribution of queueing service times \citep{HARI1968},
life testing \citep{HAAL2009}, and
the sizes of files on a computer server \citep{HOLL2006}, to name but a few.

We say a random variable $X$ follows a Lomax distribution with  scale parameter $\sigma>0$ and shape parameter $\beta>0$, if its cumulative distribution function (CDF) is
$$
F_{\sigma,\beta}(x)=P(X \leq x)=1-\left[1+\left(\frac{x}{\sigma}\right)\right]^{-\beta}, \quad x > 0,
$$
with probability density function (PDF) 
$$
f_{\sigma,\beta}(x)=\frac{\beta}{\sigma}\left(1+\frac{x}{\sigma}\right)^{-(\beta+1)}, \quad x > 0.
$$

In this paper, we investigate different approaches used for estimating the parameters of the Lomax distribution. We also draw parallels with the methods used to estimate the parameters of other Pareto distributions in order to ascertain how one might go about obtaining reliable estimators. For example, when considering the estimation of the parameters of the GPD, a number of approaches such as maximum likelihood estimation (MLE), method of moments estimation (MME), and probability weighted moments estimation (PWME) have been studied. However, not all approaches are equally viable in all settings. It is noted in \cite{HOWA1987}, for example, that the MLE experiences difficulty when estimating the parameters of the GPD under a variety of configurations of the shape and scale parameter values.
Similarly, \cite{BEKO2010} find that the usefulness of the MLE is restricted to very large sample sizes; the use of 
PWME is advocated in case of moderate sample sizes. 
For the Lomax distribution in particular, similar studies have been conducted in, for example, 
\cite{GILE2011} and \cite{SHAKE2017}.
These papers are concerned with the estimation of Lomax parameters using diverse traditional methods of estimation including MLE, MME, and PWME. The results of these methods are compared to one another and recommendations are made concerning the use of these more traditional estimation approaches. It is also found in these papers that MLE approaches are not particularly reliable when used with moderately sized samples. The conclusion from these studies is that these traditional approaches have utility in certain settings, but not all.

{The aim of this paper is twofold. First, we provide an overview of estimation techniques used for the Lomax distribution and we compare the performance of the various estimators. Second, we focus on the minimum distance estimators (MDEs) which represents an unexplored avenue for the estimation of the parameters of the Lomax distribution. While some of these procedures have already been examined in \cite{JUSC2004} for the GPD, focusing on the Lomax distribution will provide specific insight into the usefulness of these methods for this distribution. In addition, the MDEs we consider differ substantially from those proposed in \cite{JUSC2004} and \cite{BHHJ1998}, in that they only consider density power divergence measures whereas we additionally explore some distribution function based methods as well.}



Generally speaking, MDEs measure the difference between an empirical estimate of the probability density or distribution function and the theoretical function \citep{BSPA2011}. 
The use of minimum distance estimation measures has been advocated by a number of authors when modelling data containing extreme values including \cite{BOOS1981}, \cite{PASC1980} as well as \cite{PADE1981} since these estimators have good robustness qualities.

This paper studies the general case where both the scale ($\sigma$) and shape ($\beta$) parameters of the Lomax distribution are unknown. We start by considering the myriad different traditional estimation methods proposed for related distributions, including the $L$-moment estimator, the PWME, the MLE, and the MME. We then go on to propose the use of MDEs in the hope that these will be competitive alternatives to the more traditional estimators. The performance of the estimators are compared to one another via a comprehensive numerical study involving Monte Carlo simulations where the finite sample variance, relative bias, and mean squared error (MSE) of each estimator is approximated. In addition, an omnibus measure allowing one to gauge the MSE of the estimation of both $\sigma$ and $\beta$ simultaneously is employed in the simulation study.

The remainder of the article is organized as follows. In Section 2, the different parameter estimation methods are introduced and discussed. Thereafter, the Monte Carlo study is presented in Section 3, including a detailed discussion of the results of the Monte Carlo study. In Section 4, we apply the different estimators considered to a real world example. Finally, in Section 5, we conclude the paper by making a number of general observations regarding the estimators, provide some recommendations regarding the use of each.


\section{Estimation of parameters}

In this section, we discuss the parameter estimation methods which are used to estimate the shape and scale parameters of the Lomax distribution. Included here are traditional methods of estimation for parameters, along with methods based on optimising a variety of distance measures. 
{In addition to the performance of the MDEs, we provide an overview of the performance of a wide variety of competing estimators. This facilitates recommendations regarding the choice of estimator to use in various settings.} The theoretical detail relevant to the estimation of the parameters is provided in each case.


For the remainder of the section, we assume that we have data $X_1, X_2,\dots, X_n$ which is independently and identically distributed (i.i.d.) from the Lomax distribution with parameters $\beta > 0 $ and $\sigma > 0$. The order statistics based on this sample are denoted using $X_{1:n} \leq X_{2:n} \leq \cdots \leq X_{n:n}$.


\subsection{Method of moment estimators (MMEs)}
\label{sec:MME}
The MME is one of the most widely used estimation techniques. It has been used to investigate key aspects of probability distributions, such as central tendency, spread, skewness, and kurtosis. However, due to the fact that, for the Lomax distribution with shape parameter $\beta$, only moments lower than $\beta$ are finite, this approach is often not ideal for the parameters of the Lomax distribution. It is ultimately presented here and in the Monte Carlo in Section \ref{sec:MC} to illustrate its functionality for a range of values of $\beta$.

To derive the MMEs for the Lomax distribution, we note that the  $r^\textrm{th}$ theoretical raw moment of the distribution is
\begin{equation*}
\text{E}(X^r) = \mu_r =\frac{\sigma^r\Gamma(\beta-r)\Gamma(1+r)}{\Gamma(\beta)},  \quad 0 < r < \beta \quad r = 1,2,\dots.
\end{equation*}
The first and second theoretical raw moments 
are then given by 
\begin{equation}
    \label{eq:EXEX2}
    \text{E}(X)= \mu_1 =  \frac{\sigma}{\beta-1},
    \text{\quad and \quad } 
    \text{E}(X^2)= \mu_2 = \frac{2\sigma^2}{(\beta-1)(\beta-2)}.
\end{equation}
From \eqref{eq:EXEX2} we find that we can express $\beta$ and $\sigma$ as
\begin{equation}
    \label{eq:betasigMME}
    \beta = \frac{2\mu_2-2\mu_1^2}{\mu_2-2\mu_1^2} 
    \text{\quad and \quad}  
    \sigma=\mu_1(\beta-1).
\end{equation}
By simply substituting $\mu_1$ and $\mu_2$ with their sample estimators in \eqref{eq:betasigMME}, the MMEs of these parameters are obtained:
\begin{equation*}
    \label{eq:MME}
    \widehat\beta_{MM} = \frac{2\widehat\mu_2-2\widehat\mu_1^2}{\widehat\mu_2-2\widehat\mu_1^2} 
    \text{\quad and \quad}  
    \widehat\sigma_{MM}=\widehat\mu_1(\widehat\beta_{MM}-1),
\end{equation*}
where $\widehat\mu_r=\frac{1}{n}\sum_{i=1}^n X_i^r, \quad r=1,2$.

\subsection{\texorpdfstring{$L$}--moment estimators (LMEs)}
$L$-moments provide an alternative way to describe the shapes of probability distributions and are defined to be the expected values of linear combinations of order statistics. They were proposed in \cite{HOSK1990} as robust alternatives to classical moments. The method of $L$-moments has its advantages over classical moments including unbiasedness, robustness to the presence of outliers in the data and being less sensitive to sampling variability especially with data having heavy tails (which is the case in the current context). If the mean of the distribution exists, it follows that all of the $L$-moments exist and uniquely define the distribution \citep{HOSK1990}.
The $r^{\textrm{th}}$ population $L$-moment is given by:
$$
    \lambda_r = \frac{1}{r}\sum_{k=0}^{r-1}(-1)^k\binom{r-1}{k}  \text{E}\left(X_{r-k:r}\right), \quad r=1,2,\dots,
$$
where $r$ is the integer order of the $L$-moment, and $\text{E}(X_{r-k:r})$ is the expectation of the $(r-k)^{\textrm{th}}$ order statistic of a sample of size $r$. The sample $L$-moment is given by
$$
    l_r = {\binom{n}{r}}^{-1} \mathop{\sum\sum\dots\sum}_{1\leq i_1 < \ldots< i_r\leq n}\frac{1}{r} \sum_{k=0}^{r-1}(-1)^k
    \binom{r-1}{k}  X_{i_{r-k}:n}
    \quad r=1,2,\cdots, n. 
$$
The expressions of the first two sample $L$-moments are therefore
$$
l_1 = \frac{1}{n} \sum_{i=1}^n X_{i:n} 
\text{\quad and \quad}
l_2 = \frac{1}{2} {\binom{n}{2}}^{-1} \mathop{ \sum\sum}_{i>j} (X_{i:n}- X_{j:n}).
$$
Similar to the method of moments, the $L$-moments can be used to obtain parameter estimates by equating the theoretical $L$-moments of the distribution to the corresponding $L$-moments of a sample.
The $L$-moments estimators of the parameters of the Lomax distribution are therefore given by 
$$
\widehat\beta_{LM} = \frac{l_2}{2l_2 - l_1}
\text{\quad and \quad} 
\widehat\sigma_{LM} = \frac{l_1^2-l_1l_2}{2l_2-l_1},
$$
see \cite {SHAKE2017}.
$L$-moments can be a good starting point for the iterative numerical procedure needed to obtain maximum likelihood estimates \citep{HOSK1985}.

\subsection{Probability weighted moment estimators (PWMEs)}

PWMEs were proposed in \cite{GREEN1979} as a tool for estimating the parameters of probability distributions. 
Probability weighted moments of a random variable $X$ with CDF denoted $F(\cdotp)$ are quantities defined as:
$$ 
M_{p,u,v}=E\{X^p[F(X)]^u[1-F(X)]^v\}, 
$$ 
where $p$, $u$, and $v$ are integers.
Useful special cases of PWMEs are when $u=v=0$, yielding the conventional non-central moments, and the cases when $p=1$, with either $u$ or $v$ set to zero, yielding 
$$
    M_{1,u,0}=E\{X[F(X)]^u\} 
\quad\text{and}\quad
    M_{1,0,v}= E\{X[1-F(X)]^v\}, \quad u,v = 0,1,\dots .
$$
These quantities permit the following unbiased  estimators \citep[see][]{ELA2001,SHAKE2017}:
$
\widehat{M}_{1,0,0}= 
\frac{1}{n} \sum_{j=1}^n X_{j:n}, 
$
$$
\widehat{M}_{1,0,v}= 
\frac{1}{n} \sum_{j=1}^n \frac{(n-j)(n-j-1)\cdots(n-j-v+1)}{(n-1)(n-2)\cdots(n-v)}X_{j:n}, \quad v=1,2,\dots,
$$ 
and 
$$
\widehat{M}_{1,u,0}=
\frac{1}{n} \sum_{j=1}^n \frac{(j-1)(j-2)\cdots(j-u)}{(n-1)(n-2)\cdots(n-u)} X_{j:n}, \quad u=1,2,\dots .
$$

Finally, as with $L$-moment and conventional moment estimators, the PWMEs are obtained by equating the expressions for $M_{p,u,v}$ to their estimators and solving the resulting equations. In this case, the moments $M_{1,0,0}$ and $M_{1,0,1}$ are used to achieve the results
$$\widehat\beta_{PW} = \frac{2\widehat{M}_{1,0,1}-\widehat{M}_{1,0,0}}{4\widehat{M}_{1,0,1}-\widehat{M}_{1,0,0}} \ \ \ \textrm{and} \ \ \ \widehat\sigma_{PW} = \frac{2\widehat{M}_{1,0,0}\widehat{M}_{1,0,1}}{\widehat{M}_{1,0,0}-4\widehat{M}_{1,0,1}}.$$

For small samples, both 
{PWMEs} and $L$-moments estimators perform well. Indeed
because these quantities are linear combinations of each other, the calculation of one implies the calculation of the other, and thus inferences based on either are identical \citep{HOSK1990,ASQU2011}.

\subsection{Maximum likelihood estimators (MLEs)} 
Maximum likelihood estimation is one of the most popular methods for estimating the unknown parameters of probability distributions. The popularity of MLEs can be attributable to its desirable asymptotic properties such as unbiasedness and consistency. However, for small sample sizes, these properties may not hold true, resulting in biased MLEs.

In this section, we first use the MLEs to estimate the unknown parameters, and then, in the section that follows, we consider a bias-adjusted approach which reduces the bias of the MLEs to order $O(n^{-1})$, see \cite{GILE2011}. 


\subsubsection{MLE of the Lomax distribution}
The MLEs for the Lomax distribution are obtained by maximising the log-likelihood function 
\begin{equation}
    \label{eq:loglik}
    \ell:=\ell(\beta,\sigma; X_1,\dots,X_n)= n\log(\beta)-n\log(\sigma)-(1+\beta)\sum_{i=1}^n \log\left(1+\frac{X_i}{\sigma}\right),
\end{equation}
Differentiating the log-likelihood with respect to $\beta$ and $\sigma$, respectively, yields the equations:
\begin{equation}
\label{eq:betaMLE}
\frac{\partial \ell}{\partial \beta} = \frac{n}{\beta}-\sum_{i=1}^n \log\left(1+\frac{X_i}{\sigma}\right),
\end{equation}
\begin{equation}
\label{eq:sigmaMLE}
\frac{\partial \ell}{\partial \sigma} = -\frac{n}{\sigma} +\frac{1+\beta}{\sigma}\sum_{i=1}^n \frac{X_i}{\sigma + X_i}.
\end{equation}
Unfortunately, one cannot obtain closed-form solutions for both estimators using \eqref{eq:betaMLE} and \eqref{eq:sigmaMLE} so numerical procedures will be required for optimisation purposes. To simplify the optimisation problem, the expressions are reduced to the optimisation of a single parameter 
{(this is the approach followed in \cite{Renext} using their \texttt{flomax} function).} This is accomplished by first noting that when setting \eqref{eq:betaMLE} to zero and solving we get the MLE, $\widehat\beta_{ML}$, expressed in terms of $\widehat\sigma_{ML}$
\begin{equation}
\label{eq:betahatMLE}
\widehat\beta_{ML} = \frac{n}{\sum_{i=1}^n \log \left(1+{X_i}/{\widehat\sigma_{ML}}\right)}.
\end{equation}
The maximised log-likelihood, $\widehat\ell:=\ell(\widehat\beta_{ML},\widehat\sigma_{ML}; X_1,\dots,X_n)$, can therefore be expressed solely in terms of the MLE, $\widehat\sigma_{ML}$, by substituting the form of $\widehat\beta_{ML}$ given in \eqref{eq:betahatMLE} and $\widehat\sigma_{ML}$ into \eqref{eq:loglik}:
\begin{align*}
\widehat\ell &=-n\log\left(\frac{1}{n}\sum_{i=1}^n\log\left(1+\frac{X_i}{\widehat\sigma_{ML}}\right)\right)-
n\log(\widehat\sigma_{ML})-
n\left(\frac{1}{n}\sum_{i=1}^n \log \left(1+\frac{X_i}{\widehat\sigma_{ML}} \right)+1\right).
\end{align*}
Then, using R's one-dimensional optimisation function, \texttt{optimize} \citep{R2021}, the MLEs $\widehat\sigma_{ML}$ and $\widehat\beta_{ML}$ (via \eqref{eq:betahatMLE}) are obtained.

\subsubsection{MLEs of the Lomax distribution adjusted for bias (MLE.b)}
For small sample sizes, the MLEs sometimes perform poorly. In order to solve this problem, \cite{GILE2011} proposed the use of second-order bias-adjusted versions of the MLEs. To obtain these estimators the bias of the MLE is first approximated and then subtracted from the MLE to produce the bias-corrected estimate.

Let $K$ be the Fisher information matrix for the Lomax distribution. It can be shown that
$$
K^{-1} = \left[
\begin{array}{cc}
\frac{\sigma^2}{n\beta}(\beta+2)(\beta+1)^2 & \sigma\beta(\beta+1)(\beta+2)/n\\
\sigma\beta(\beta+1)(\beta+2)/n & \beta^2(\beta+1)^2/n
\end{array}
\right].
$$
{The $O(n^{-1})$ bias is obtained in \cite{GILE2011} as}
$$Bias\binom{\widehat\sigma_{ML}}{\widehat\beta_{ML}}={K}^{-1}{A}\text{vec}({K}^{-1}) + O(n^{-2}),
$$
where $\text{vec}(\cdotp)$ is an operator that, when applied to a $(m \times n)$ matrix, produces a single column vector of dimension $(mn \times 1)$ by piling up the column vectors below one another, and the matrix $A$ is given by
$$
 A= n \left[
\begin{array}{cccc}
2\beta/\sigma^3(\beta+2)(\beta+3) & -1/\sigma^2(\beta+1)(\beta+2) & \beta/\sigma^2(\beta+2)^2 & -1/\sigma(\beta+1)^2\\
-1/\sigma^2(\beta+1)(\beta+2) & 0 & -1/\sigma(\beta+1)^2 & 1/\beta^3
\end{array}
\right].
$$
The bias-adjusted estimators 
{proposed in \cite{GILE2011}} are then obtained by subtracting an estimator of this approximated bias from the original MLEs:
\begin{equation}
\label{eq:biasadj}
\binom{
{\widehat{\sigma}_{MLB}}}{
{\widehat{\beta}_{MLB}}} = \binom{\widehat\sigma_{ML}}{\widehat\beta_{ML}}-\widehat{Bias}\binom{\widehat\sigma_{ML}}{\widehat\beta_{ML}} = \binom{\widehat\sigma_{ML}}{\widehat\beta_{ML}}- \widehat{K}^{-1}\widehat{A}\text{vec}(\widehat{K}^{-1}),
\end{equation}
where $$\widehat{K} = K|_{\sigma=\widehat\sigma_{ML};\beta=\widehat\beta_{ML}} \quad\text{and}\quad \widehat{A} = A|_{\sigma=\widehat\sigma_{ML};\beta=\widehat\beta_{ML}}.$$
The bias-adjusted estimators in \eqref{eq:biasadj} have simple closed-form expressions, making them  computationally appealing. It is expected that the bias adjusted-estimators, 
{$\widehat{\sigma}_{MLB}$} and 
{$\widehat{\beta}_{MLB}$} will outperform $\widehat\sigma_{ML}$ and $\widehat\beta_{ML}$, respectively for smaller to moderate samples.

\subsection{Minimum distance estimators (MDEs)}

MDEs are estimators minimizing, for example, some distance measure between the theoretical CDF and the empirical distribution function (EDF) of the observed sample data. Note that we define the EDF 
as
 $$F_{n}(x)=\frac{1}{n}\sum_{i=1}^{n} I \left(X_{i:n} \leq x\right),$$
where $I(\cdot)$ is the indicator function and $n$ is the sample size.
The main idea is to see how closely the theoretical CDF is to the EDF and the distance between two probability distributions captures the difference in information between them. Minimum distance estimation was first subjected to the in-depth study in a series of papers discussed in \cite{WOLF1953} and has since been considered as a method for deriving efficient and robust estimators by \cite{BERA1977}, \cite{BERA1978}, \cite{BOOS1981} as well as \cite{PASC1980}, among others.

Several minimum distance estimation methods have been proposed including those derived from empirical distribution function such as Cram\'{e}r-von Mises and those measures derived from entropy such as Kullback-Liebler divergence. In this section we explore several distance measures; the Cram\'{e}r-von Mises (CVM) statistic, least squares, and some phi-divergence measures such as Kullback-Leibler(KL) divergence, total variation (TV) distance and chi-square divergence. 

{In Section \ref{ress} we calculate these MDEs using the \texttt{optim} function in R, specifically we use the ``BFGS'' optimisation method; see \cite{Bro1970}, \cite{Fle1970}, \cite{Gol1970} and \cite{Sha1970}. This is a quasi-Newton method reliant on gradients to perform optimisation. As initial values we use the LMEs as they are available in closed-form and the LMEs generally perform well compared to other estimators.}

\subsubsection{A Cram\'{e}r-von Mises distance measure (MDE.CvM)}

The Cram\'{e}r-von Mises (CVM) statistic is based on the squared integral difference between the EDF and the theoretical distribution. It has the form
$$
    Q_n = n \int_{-\infty}^{\infty}\left[F_n(x) - F(x)\right]^2 dF(x).
$$ 

We can define the CVM \emph{distance measure} used to estimate the parameters of the Lomax distribution using the MDE approach as follows:
$$
    D_n^{CM}(\sigma,\beta) = \int_{0}^{\infty} \left[ F_n(x) - F_{\sigma,\beta}(x)  \right]^2 f_{\sigma,\beta}(x)dx, 
$$
where $F_{\sigma,\beta}(x)$ and $f_{\sigma,\beta}(x)$ are the CDF and PDF of the Lomax distribution with scale parameter $\sigma$ and shape parameter $\beta$, respectively. This distance measure permits a simple calculable form given by
\begin{align*}
    D^{CM}_n(\sigma,\beta) &= \frac{1}{12n} + \sum_{j=1}^n \left( F_{\sigma,\beta}\left(X_{j:n}\right)-\frac{2j-1}{2n} \right)^2  \\ 
 &= \frac{1}{12n}+\sum_{j=1}^n  \left(1-\left(\frac{\sigma}{X_{j:n}+\sigma}\right)^\beta-\frac{2j-1}{2n}\right)^2.
 \end{align*}
Finally, the resulting estimators for $\sigma$ and $\beta$ can be expressed as simply the values of those parameters that minimise this distance measure:
$$
  (\widehat\sigma_{CM},\widehat\beta_{CM}) =\arg\min_{(\sigma,\beta)}D_n^{CM}(\sigma,\beta).
$$
 
\subsubsection{A `squared difference' distance estimator (MDE.SD)}
An alternative to the Cram\'{e}r-von Mises distance measure is the much simpler `squared difference' measure. This distance measure focuses only on the squared difference between the quantities $F_n$ and $F$ and has the following general expression:
$$
    S_n = \int_{-\infty}^{\infty}\left[F_n(x) - F(x)\right]^2
    dF_n(x).
$$
Now, using of the Lomax distribution function, $F_{\sigma,\beta}(\cdotp)$, the distance measure can be defined as
$$
    D_n^{SD}(\sigma,\beta) = \int_{0}^{\infty}\left[F_n(x) - F_{\sigma,\beta}(x)\right]^2
    dF_n(x),
$$
where, upon simplification, we obtain the following tractable calculation form:
\begin{align*}
  D_n^{SD}(\sigma,\beta) &= \sum_{j=1}^n \left( F_{\sigma,\beta}(X_{j:n}) - \frac{j}{n+1}\right)^2\\
 &= \sum_{j=1}^n \left(1-\left(\frac{\sigma}{X_{j:n}+\sigma}\right)^{\beta}-\frac{j}{n+1}\right)^2.
 \end{align*}
Therefore, the estimators obtained by minimising this `squared difference' distance measure can be expressed as 
$$
(\widehat\sigma_{SD}, \widehat\beta_{SD})= \arg\min_{(\sigma,\beta)}D_n^{SD}(\sigma,\beta).
$$
\subsubsection{The \texorpdfstring{$\phi$-}-divergence distance measures}
The $\phi$-divergence distance measures represent a broad class of distance measures that describe the distance between two densities $f$ and $g$, and is defined as:
\begin{equation}\label{eq:deltaphi}
    \delta(f,g) = \text{E}\left[ \phi \left(\frac{f(X)}{g(X)}\right)\frac{g(X)}{f(X)}\right],
\end{equation}
where $X$ is a random variable from a distribution function with density $f$ and $\phi(\cdot)$ is a convex function such that $\phi(1)= 0$ and $\left.\frac{\partial^2 \phi(x)}{\partial x^2 }\right|_{x=1}=\phi''(1) > 0.$
In \eqref{eq:deltaphi}, we set $g(x)$ to be the Lomax density function, $f_{\sigma,\beta}(x)$,  and  set $f(x)$ to be some empirical estimator for the density, $\widehat{f}(x)$, to get the 
following distance measure which is to be minimised in order to estimate the parameters $\sigma$ and $\beta$:
$$
    D_n^{\phi}(\sigma,\beta) = 
    \frac{1}{n}\sum_{j=1}^n \phi \left(\frac{\widehat{f} (X_{j})}{f_{\sigma,\beta}(X_j)}\right)\frac{f_{\sigma,\beta}(X_{j})}{\widehat{f}(X_{j})}.
$$
The required parameters estimates are
$$
    (\widehat{\sigma},\widehat{\beta}) = \arg\min_{(\sigma,\beta)}D_n^\phi (\sigma,\beta).
$$
In this setting, we estimate the density using the kernel density estimator, $\widehat{f}_h(x)$, defined as
$$
    \widehat{f}_h(x) = \frac{1}{nh}\sum_{i=1}^n k \left( \frac{x-X_i}{h}\right),
$$
where $k(\cdot)$ is the kernel function, chosen to be the standard normal density function, and $h$ is the bandwidth, chosen using Silverman's rule-of-thumb $h=0.9n^{-1/5}\min(s,{(q_3-q_1)}/{1.34})$, where $q_1$ and $q_3$ denote the sample quartiles and $s$ denotes the sample standard deviation \citep{SIL1986}. The practical implementation of the kernel density estimator is executed using the \texttt{density} function in R \citep{R2021}.


Different choices of the function $\phi(\cdotp)$ lead to different forms of this distance measure. We consider three choices of $\phi(\cdot)$, these choices are described below.

\paragraph{Kullback-Liebler $\phi$-divergence distance measure (MDE.KL)}

The Kullback-Liebler $\phi$-divergence distance measure can be obtained by setting
$$
    \phi(t) = t\log(t).
$$ 
The resulting distance measure is
$$
    D_n^{KL}(\sigma,\beta) = 
    \frac{1}{n}\sum_{j=1}^n 
    \log \left(\frac{\widehat{f}_h (X_j)}{f_{\sigma,\beta}(X_j)}\right),
$$
and we denote the estimator by $(\widehat\sigma_{KL}, \widehat\beta_{KL})$. 
\cite{BHHJ1998} note that this  Kullback-Liebler form simply represents a robust extension of MLE. This approach is also followed in \cite{JUSC2004} where a general class of density power divergent estimators are employed to estimate the  parameters of the GPD.  

\paragraph{Chi-squared $\phi$-divergence distance measure (MDE.$\chi^2$)}
Setting $\phi(t) = (t-1)^2$ yields the chi-square $\phi$-divergence distance measure, denoted   
$$
 D_n^{CS}(\sigma,\beta)= 
\frac{1}{n}\sum_{j=1}^n  
\frac{\left(\widehat{f}_h(X_{j})-f_{\sigma,\beta}(X_j)\right)^2}
{f_{\sigma,\beta}(X_{j})\widehat{f}_h(X_{j})},
$$
and we express the resulting estimator as $(\widehat\sigma_{CS}, \widehat\beta_{CS})$ 

\paragraph{Total variation $\phi$-divergence distance measure (MDE.TV)} When specifying
$\phi(t) = |t-1|$, we obtain the total variation $\phi$-divergence distance measure, given by 
$$
    D_n^{TV}(\sigma,\beta)=
    \frac{1}{n}\sum_{j=1}^n \left| 1 - \frac{f_{\sigma,\beta}(X_{j})}{\widehat{f}_h(X_{j})} \right|,
$$
and we designate the resulting estimator by $(\widehat\sigma_{TV}, \widehat\beta_{TV})$.





\section{Finite sample results}\label{ress}

Below we consider the Monte Carlo setting used before turning our attention to the results obtained.

\subsection{Monte Carlo simulation settings}
\label{sec:MC}

In order to compare the performance of the various estimators to one another, a comprehensive Monte Carlo simulation study was conducted. The simulation study sets out to approximate several distributional properties of the estimators, including the expected value, variance, 
{relative bias (RB)} and mean squared error (MSE). Using these various approximations, it will be possible to compare the estimators to one another in a sensible manner.

The Monte Carlo was conducted by simulating $MC = 10\,000$ samples of size 
{$n\in\{30,50,100,200,500\}$} from the Lomax distribution using a variety of parameter settings. For the parameter $\sigma$, the values used included 
{$\sigma\in\{0.5,1,2\}$} whereas for the parameter $\beta$, we used $\beta\in\{1.1,1.5,2,2.1\}$. 
{We consider only values of $\beta$ exceeding $1$, since the mean of the distribution is infinite if $\beta \leq 1$, see \eqref{eq:EXEX2}. Furthermore, the value of $\beta=2.1$ is included so as to consider the case where the variance exists.}
Note that, for sample size $n=30$, we omit the results pertaining to $\sigma=0.5$. The majority of the simulations in this setting produced data sets for which the optimisation procedures used fail to converge, rendering the results unusable.

For each randomly generated sample the parameters were estimated and collected to ultimately obtain the approximations of the distributional properties mentioned. All calculations were done using R \citep{R2021}.

In addition to calculating the MSE for each parameter separately, a combined MSE yielding a single value was also calculated as follows. For every pair of estimates, define
\begin{align*}
\varsigma_{j}&=\left\lVert
\begin{bmatrix} 
 \widehat\beta_{j}\mathstrut \\ 
  \widehat\sigma_{j}\mathstrut 
  \end{bmatrix}
  -
  \begin{bmatrix} 
  \beta  \\ 
  \sigma 
  \end{bmatrix}
\right\rVert^2 \\
&=\left(\widehat\beta_{j}-\beta\right)^2+\left(\widehat\sigma_{j}-\sigma\right)^2,
\end{align*}  
then we calculate 
$\textrm{TMSE}=\frac{1}{MC}\sum_{j=1}^{MC}\varsigma_{j}$ 
as the 
measure of the 
total mean squared error (TMSE)
of the estimation technique.

\begin{figure}[!h!]
\centering
\begin{tabular}{cc}
\includegraphics[width=0.5\textwidth]{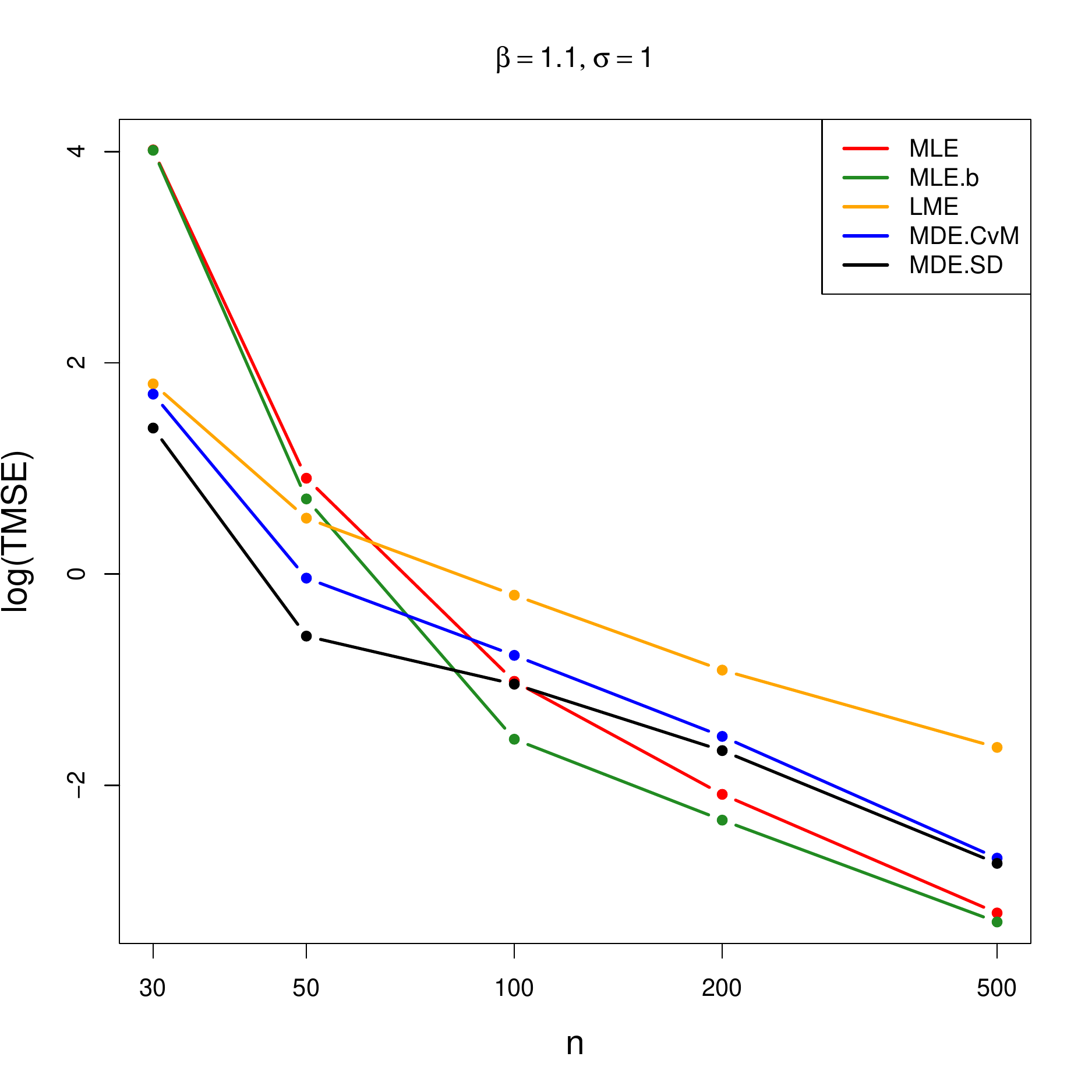} &
\includegraphics[width=0.5\textwidth]{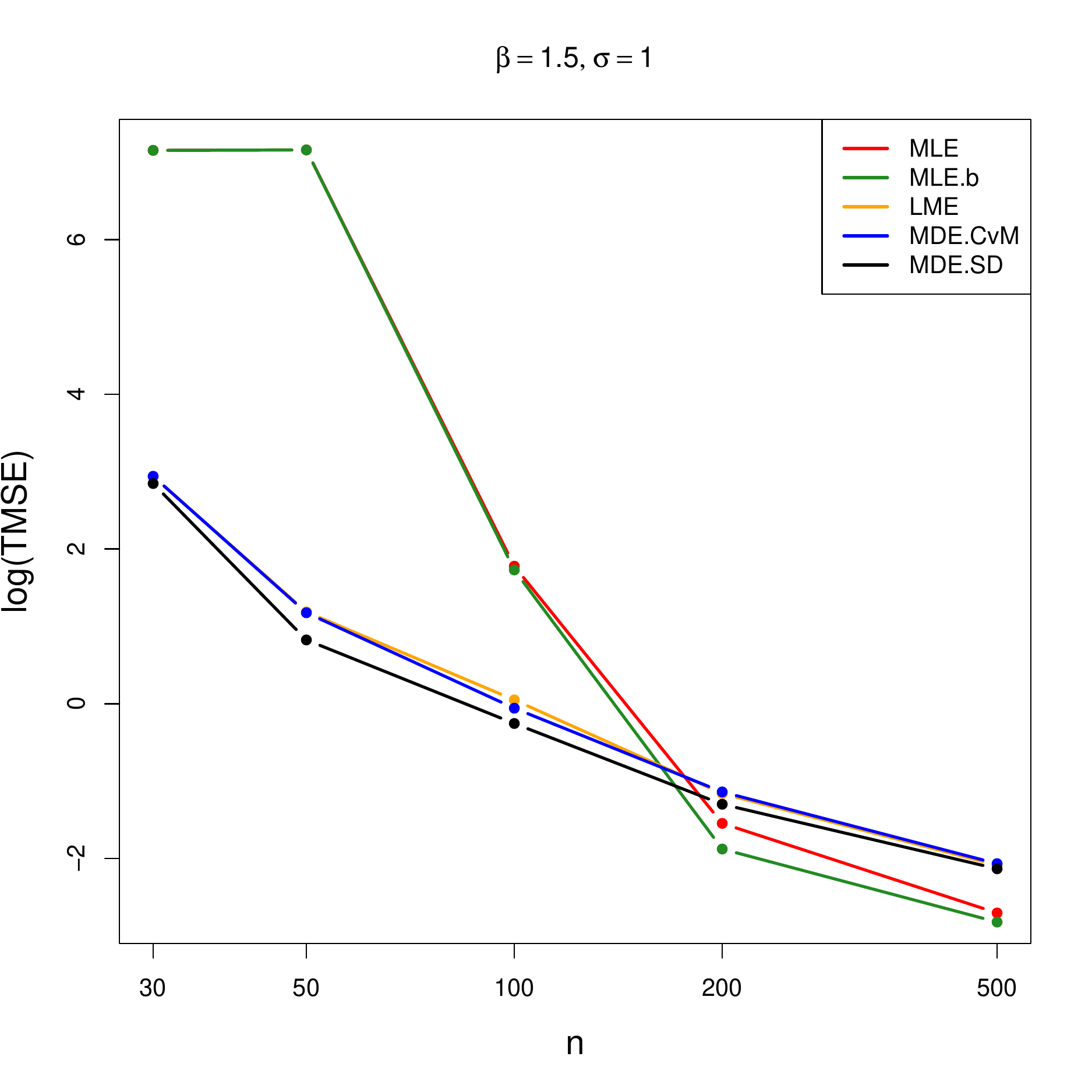} \\
\includegraphics[width=0.5\textwidth]{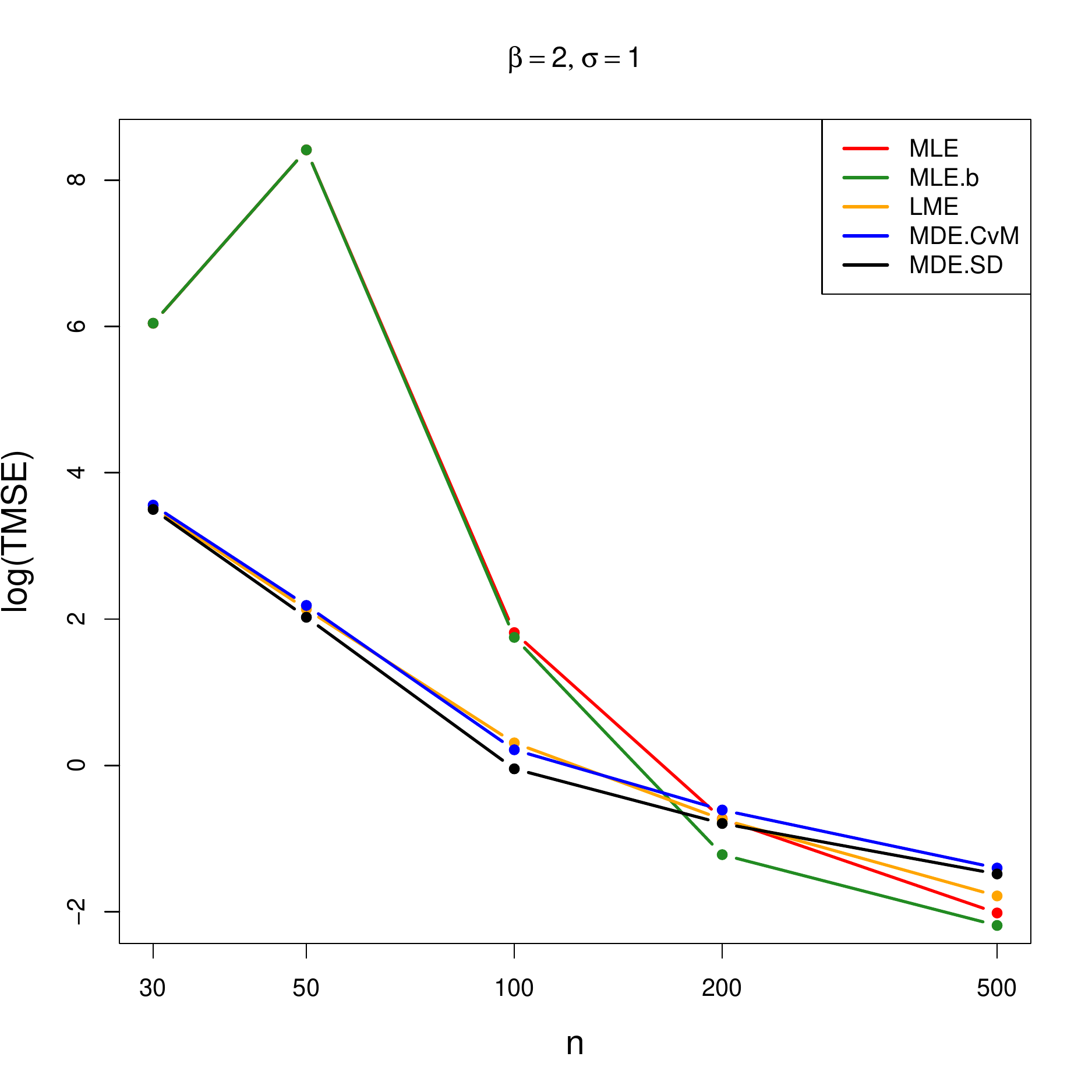} &
\includegraphics[width=0.5\textwidth]{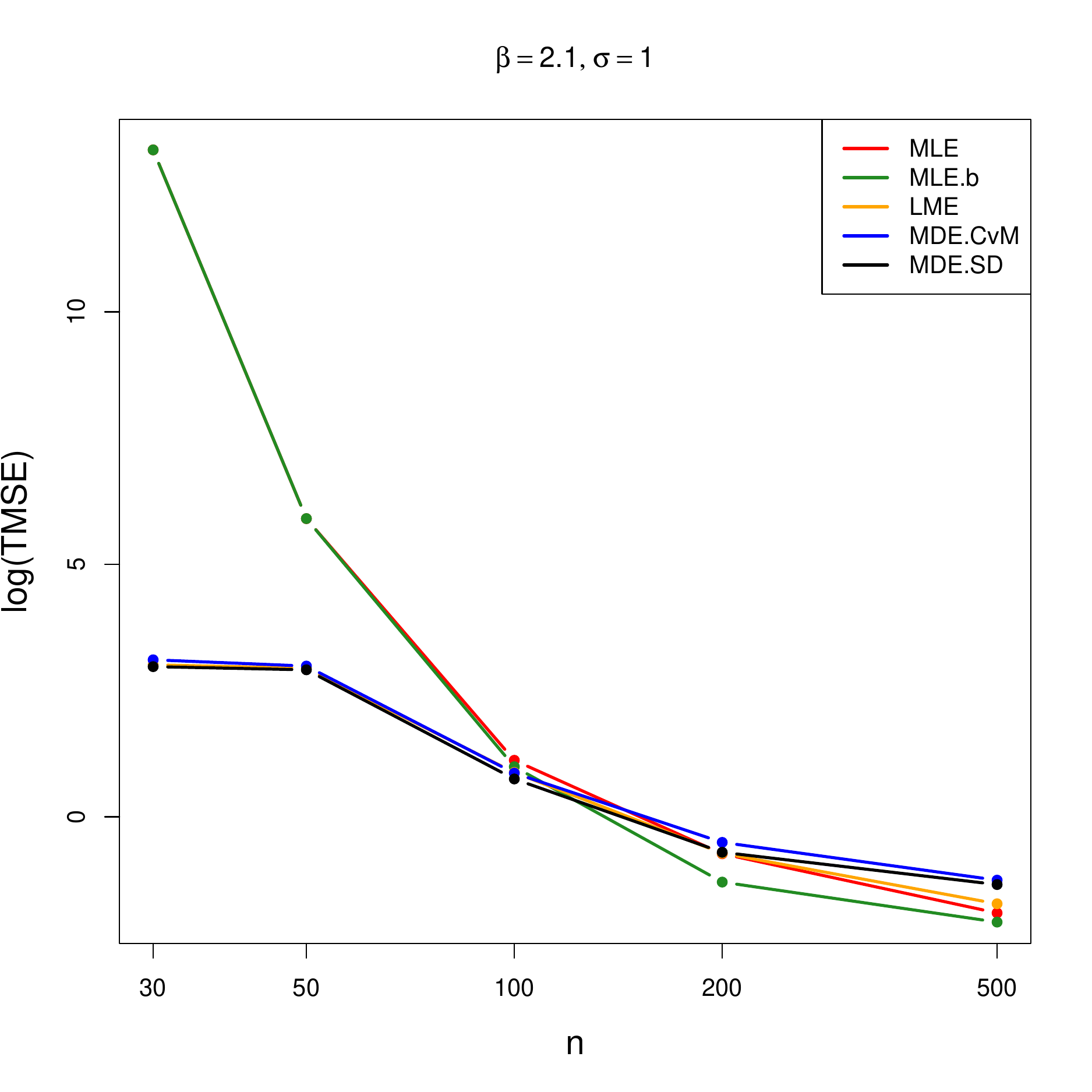} 
\end{tabular}
\caption{Plots of the log of the 
TMSE
measure against sample size for 5 different estimation techniques when \(\sigma=1\).}
\label{fig:accplots1}
\end{figure}

\begin{figure}[!h!]
\centering
\begin{tabular}{cc}
\includegraphics[width=0.5\textwidth]{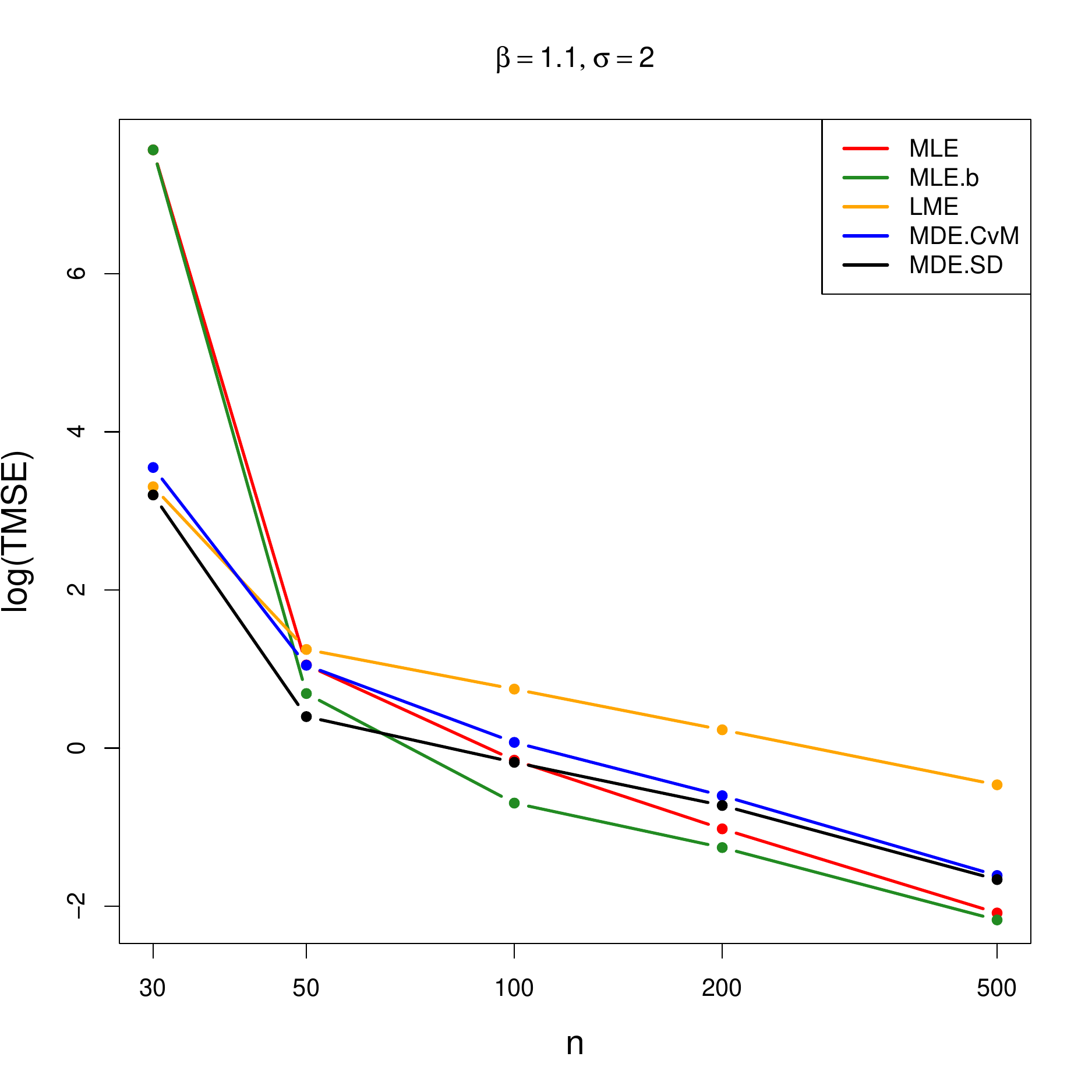} &
\includegraphics[width=0.5\textwidth]{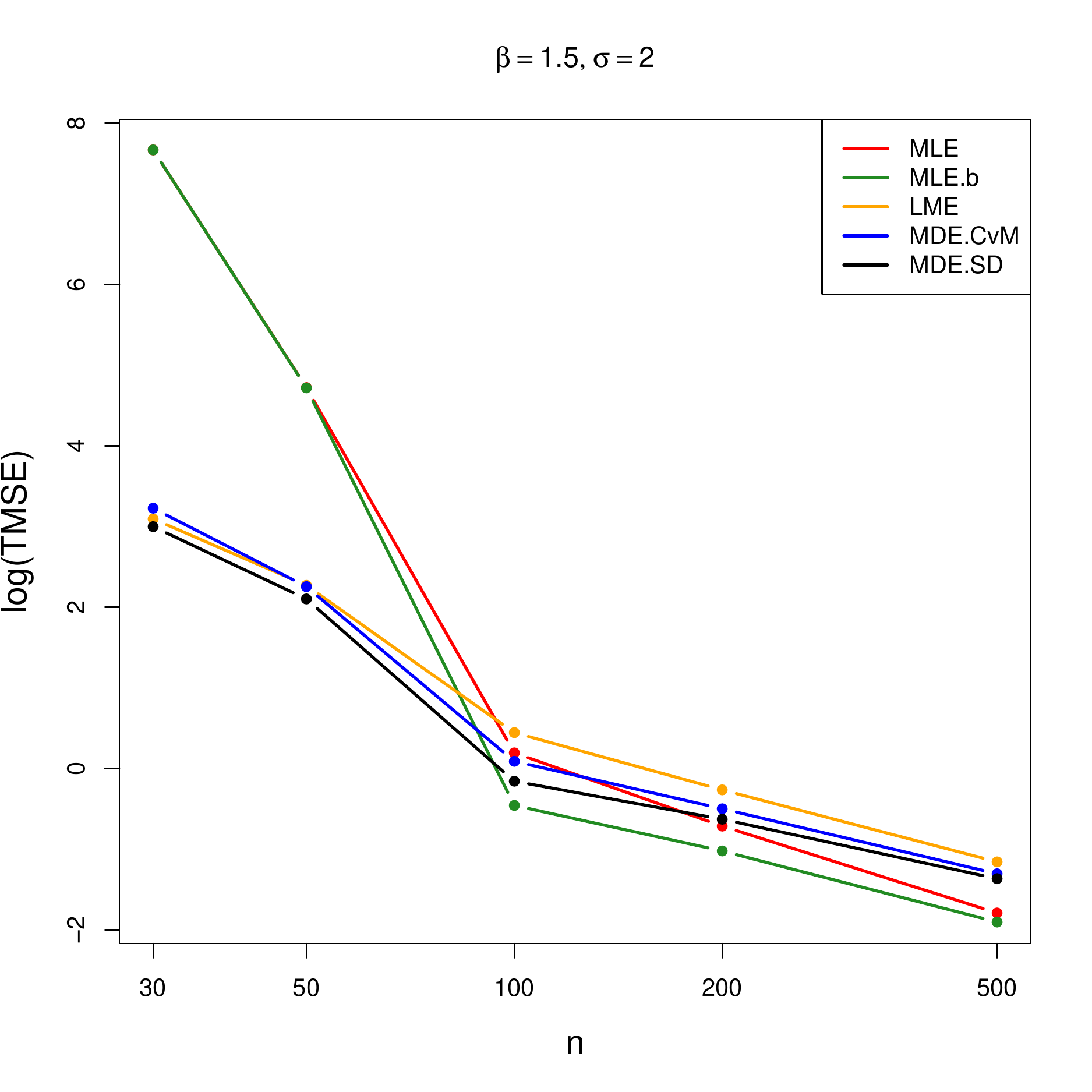} \\
\includegraphics[width=0.5\textwidth]{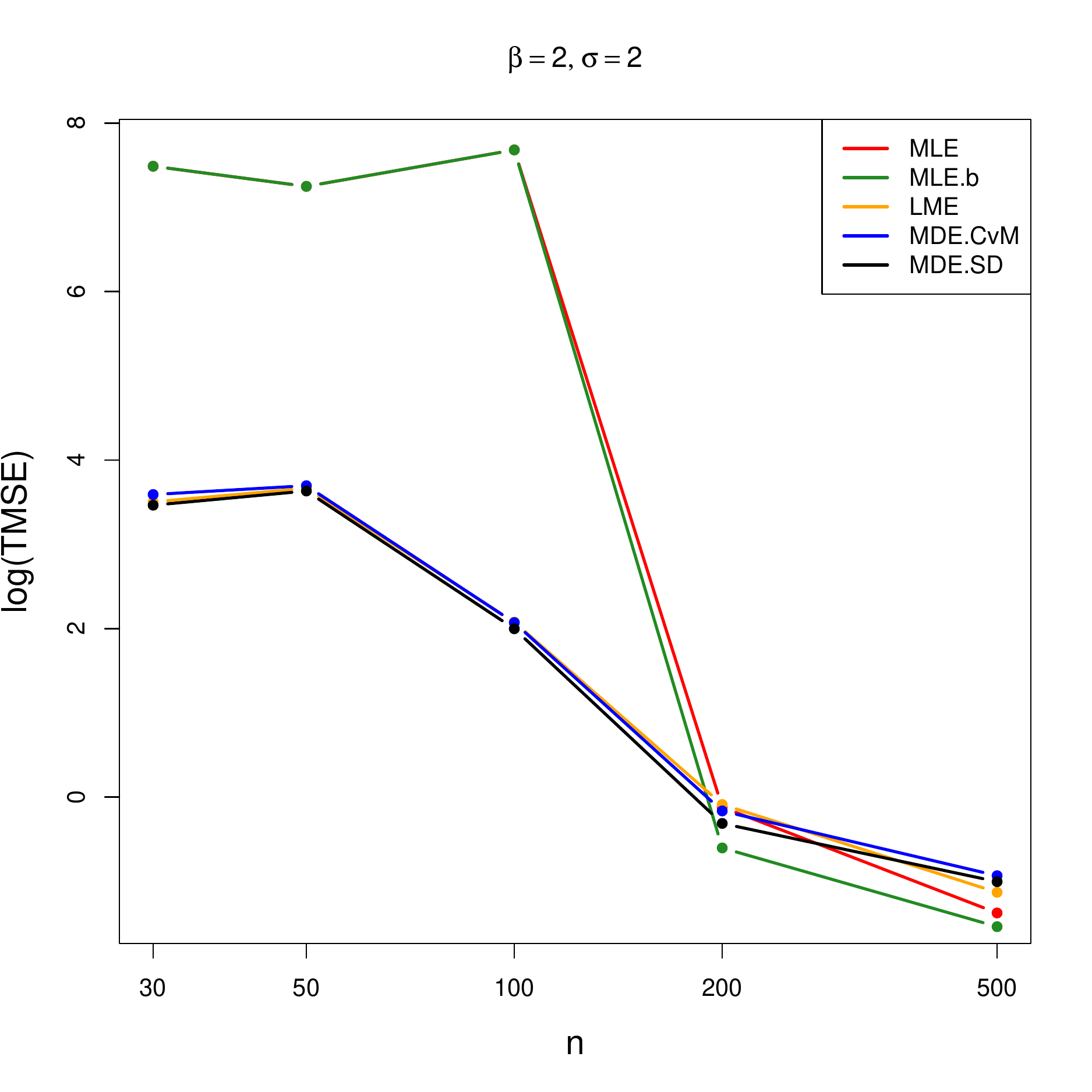} &
\includegraphics[width=0.5\textwidth]{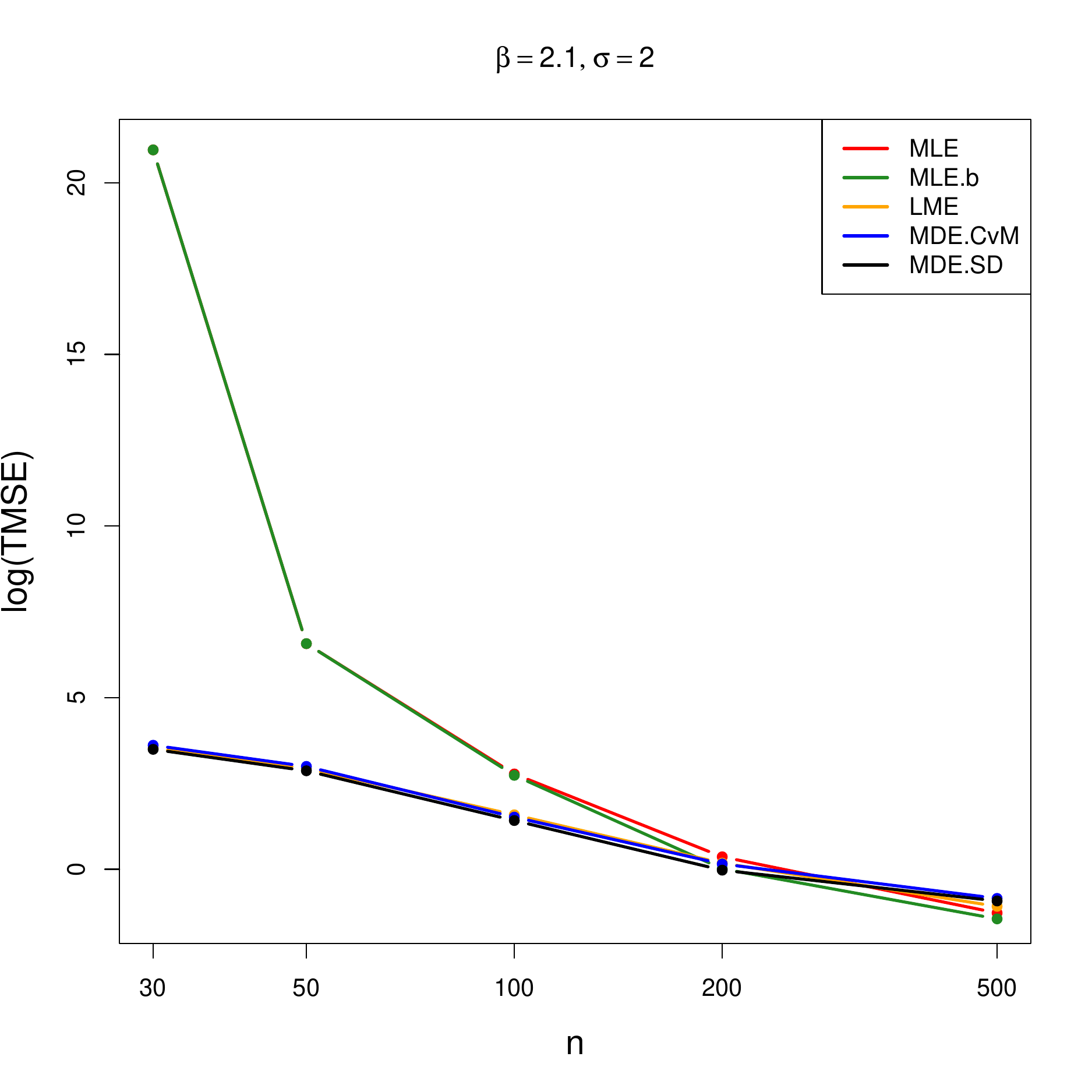} 
\end{tabular}
\caption{Plots of the log of the 
\color{red}
TMSE
\color{black}
measure against sample size for 5 different estimation techniques when \(\sigma=2\).}
\label{fig:accplots2}
\end{figure}

Tables 
\ref{tbl:n30sig1} 
to 
\ref{tbl:n100sig2} 
present the Monte Carlo approximations of the expected value, variance, 
{RB} and MSE for all estimators of $\sigma$ and $\beta$ as well as the 
TMSE
described above, for sample sizes $n=30$, $50$, and $100$. 
The remaining tables for samples sizes $n=200$ and $500$ are presented in the appendix.
{The RB of an estimator is defined to be the bias divided by the true parameter value. Figures \ref{fig:accplots1} and \ref{fig:accplots2} show the logarithm of the 
TMSE
as a function of sample size. We choose to use a logarithmic scale as this increases the separation between the lines on the graph, which makes the figures easier to interpret.}
\afterpage{%
    \clearpage%
\begin{landscape}
      \begin{table}[htbp]
  \centering
\caption{Comparison of different estimation methods for different values of $\beta$ ($\sigma=1$ and $n=30$). \label{tbl:n30sig1}}

}
\end{table}
\begin{table}[htbp]
  \centering
  \caption{Comparison of different estimation methods for different values of $\beta$ ($\sigma=2$ and $n=30$). \label{tbl:n30sig2}}
\resizebox{1.3\textwidth}{!}{
}
\end{table}

      \begin{table}[htbp]
  \centering
\caption{Comparison of different estimation methods for different values of $\beta$ ($\sigma=0.5$ and $n=50$). \label{tbl:n50sig05}}
}
\end{table}

      \begin{table}
   \caption{Comparison of different estimation methods for different values of $\beta$ ($\sigma=1$ and $n=50$). \label{tbl:1}}
 \begin{center}
  \resizebox{1.3\textwidth}{!}{
    \begin{tabular}{|cc|r|r|r|r|r|r|r|r|r|}
\cline{2-11}    \multicolumn{1}{r|}{} & \textbf{METHODS} & \multicolumn{1}{l|}{\textbf{MLE}} & \multicolumn{1}{l|}{\textbf{MLE.b}} & \multicolumn{1}{l|}{\textbf{LME}} & \multicolumn{1}{l|}{\textbf{MDE.CvM}} & \multicolumn{1}{l|}{\textbf{MDE.SD}} & \multicolumn{1}{l|}{\textbf{MDE.$\chi^2$}} & \multicolumn{1}{l|}{\textbf{MDE.TV}} &\multicolumn{1}{l|}{\textbf{MDE.KL}} & \multicolumn{1}{l|}{\textbf{MME}} \\
\cline{2-11}
      \multicolumn{11}{c}{$\beta=1.1,  \sigma=1$ $n=50$}\\
      \hline
\multirow{2}[4]{*}{Mean} & \multicolumn{1}{|c|}{$\widehat\beta$} & 1.323&1.106&1.614&1.255&1.121&0.505&1.381&1.323&2.809
\\
\cline{2-11}          & \multicolumn{1}{|c|}{$\widehat\sigma$} &
1.339&0.986&1.792&1.228&1.053&2.513&3.897&1.339&6.037
\\
\hline
\multirow{2}[4]{*}{Var} & \multicolumn{1}{|c|}{$\widehat\beta$} & 0.624&0.553&{0.219}&0.296&\textbf{0.182}&50.960&31.357&0.625&3.663
\\
\cline{2-11}          & \multicolumn{1}{|c|}{$\widehat\sigma$} &
1.688&1.483&{0.586}&0.590&\textbf{0.371}&29.810&278.536&1.692&17.261
\\
\hline
\multirow{2}[4]{*}{RB} & \multicolumn{1}{|c|}{$\widehat\beta$} & 20.264&\textbf{0.555}&46.764&14.073&1.946&-54.082&25.500&20.264&155.355
\\
\cline{2-11}          & \multicolumn{1}{|c|}{$\widehat\sigma$} &
33.920&\textbf{-1.400}&079.230&22.840&5.260&151.250&289.670&33.920&503.740
\\
\hline
\end{tabular}
 }
 \end{center}
\end{table}
      \begin{table}
   \caption{Comparison of different estimation methods for different values of $\beta$ ($\sigma=2$ and $n=50$). \label{tbl:2}}
  \begin{center}
  \resizebox{1.3\textwidth}{!}{
    \begin{tabular}{|cc|r|r|r|r|r|r|r|r|r|}
\cline{2-11}    \multicolumn{1}{r|}{} & \textbf{METHODS} & \multicolumn{1}{l|}{\textbf{MLE}} & \multicolumn{1}{l|}{\textbf{MLE.b}} & \multicolumn{1}{l|}{\textbf{LME}} & \multicolumn{1}{l|}{\textbf{MDE.CvM}} & \multicolumn{1}{l|}{\textbf{MDE.SD}} & \multicolumn{1}{l|}{\textbf{MDE.$\chi^2$}} & \multicolumn{1}{l|}{\textbf{MDE.TV}} &\multicolumn{1}{l|}{\textbf{MDE.KL}} & \multicolumn{1}{l|}{\textbf{MME}} \\
\cline{2-11}
     \multicolumn{11}{c}{$\beta=1.1,  \sigma=2$ $n=50$}\\
     \hline
\multirow{2}[4]{*}{Mean} & \multicolumn{1}{|c|}{$\widehat\beta$} & 1.301&1.083&1.591&1.259&1.128&-2.292&1.191&1.301&2.732
\\
\cline{2-11}          & \multicolumn{1}{|c|}{$\widehat\sigma$} &
2.383&1.733&3.168&2.271&1.949&4.399&5.927&2.383&10.936
\\
\hline
\multirow{2}[4]{*}{Var} & \multicolumn{1}{|c|}{$\widehat\beta$} & 0.290&0.218&0.182&0.294&\textbf{0.161}&402.879&16.815&0.290&0.613
\\
\cline{2-11}          & \multicolumn{1}{|c|}{$\widehat\sigma$} &
2.385&1.704&1.694&2.458&\textbf{1.325}&54.707&390.725&2.385&42.258
\\
\hline
\multirow{2}[4]{*}{RB} & \multicolumn{1}{|c|}{$\widehat\beta$} & 18.300&\textbf{-1.546}&44.636&14.482&2.527&-308.327&8.309&18.300&148.391
\\
\cline{2-11}          & \multicolumn{1}{|c|}{$\widehat\sigma$} &
19.170&-13.365&58.390&13.540&\textbf{-2.560}&119.940&196.370&19.170&446.820
\\
\hline
\end{tabular}
}
\end{center}
\end{table}

      \begin{table}[htbp]
  \centering
\caption{Comparison of different estimation methods for different values of $\beta$ ($\sigma=0.5$ and $n=100$). \label{tbl:n100sig05}}
}
\end{table}
      \begin{table}
   \caption{Comparison of different estimation methods for different values of $\beta$ ($\sigma=1$ and $n=100$). \label{tbl:3}}
  \begin{center}
  \resizebox{1.3\textwidth}{!}{
    \begin{tabular}{|cc|r|r|r|r|r|r|r|r|r|}
\cline{2-11}    \multicolumn{1}{r|}{} & \textbf{METHODS} & \multicolumn{1}{l|}{\textbf{MLE}} & \multicolumn{1}{l|}{\textbf{MLE.b}} & \multicolumn{1}{l|}{\textbf{LME}} & \multicolumn{1}{l|}{\textbf{MDE.CvM}} & \multicolumn{1}{l|}{\textbf{MDE.SD}} & \multicolumn{1}{l|}{\textbf{MDE.$\chi^2$}} & \multicolumn{1}{l|}{\textbf{MDE.TV}} &\multicolumn{1}{l|}{\textbf{MDE.KL}} & \multicolumn{1}{l|}{\textbf{MME}} \\
\cline{2-11}
       \multicolumn{11}{c}{$\beta=1.1,  \sigma=1$ $n=100$}\\
       \hline
 \multirow{2}[4]{*}{Mean} & \multicolumn{1}{|c|}{$\widehat\beta$} & 1.209&1.113&1.473&1.208&1.146&-0.074&0.775&1.209&2.423
 \\
 \cline{2-11}          & \multicolumn{1}{|c|}{$\widehat\sigma$} &
 1.167&1.012&1.593&1.161&1.079&1.733&1.862&1.167&5.565
 \\
\hline
\multirow{2}[4]{*}{Var} & \multicolumn{1}{|c|}{$\widehat\beta$} & 0.096&\textbf{0.065}&0.085&0.139&0.112&84.437&0.954&0.096&0.163
\\
\cline{2-11}          & \multicolumn{1}{|c|}{$\widehat\sigma$} &
0.226&\textbf{0.144}&0.244&0.287&0.232&5.603&4.037&0.226&5.525
\\
\hline
\multirow{2}[4]{*}{RB} & \multicolumn{1}{|c|}{$\widehat\beta$} & 9.909&\textbf{1.209}&33.882&9.846&4.155&-106.727&-29.573&9.909&120.300
\\
\cline{2-11}          & \multicolumn{1}{|c|}{$\widehat\sigma$} &
16.730&\textbf{1.220}&59.300&16.120&7.880&73.250&86.170&16.730&456.520
\\
\hline
\end{tabular}
}
\end{center}
\end{table}

     \begin{table}
   \caption{Comparison of different estimation methods for different values of $\beta$ ($\sigma=2$ and $n=100$). \label{tbl:n100sig2}}
  \begin{center}
  \resizebox{1.3\textwidth}{!}{
    \begin{tabular}{|cc|r|r|r|r|r|r|r|r|r|}
\cline{2-11}    \multicolumn{1}{r|}{} & \textbf{METHODS} & \multicolumn{1}{l|}{\textbf{MLE}} & \multicolumn{1}{l|}{\textbf{MLE.b}} & \multicolumn{1}{l|}{\textbf{LME}} & \multicolumn{1}{l|}{\textbf{MDE.CvM}} & \multicolumn{1}{l|}{\textbf{MDE.SD}} & \multicolumn{1}{l|}{\textbf{MDE.$\chi^2$}} & \multicolumn{1}{l|}{\textbf{MDE.TV}} &\multicolumn{1}{l|}{\textbf{MDE.KL}} & \multicolumn{1}{l|}{\textbf{MME}} \\
\cline{2-11}
    \multicolumn{11}{c}{$\beta=1.1,  \sigma=2$ $n=100$}\\
    \hline
\multirow{2}[4]{*}{Mean} & \multicolumn{1}{|c|}{$\widehat\beta$} & 1.214&1.118&1.480&1.205&1.144&-1.901&0.772&1.214&2.428
\\
\cline{2-11}          & \multicolumn{1}{|c|}{$\widehat\sigma$} &
2.244&1.947&3.038&2.213&2.059&3.420&3.532&2.244&11.122
\\
\hline
\multirow{2}[4]{*}{Var} & \multicolumn{1}{|c|}{$\widehat\beta$} & 0.078&\textbf{0.051}&0.076&0.111&0.091&303.093&0.231&0.078&0.139
\\
\cline{2-11}          & \multicolumn{1}{|c|}{$\widehat\sigma$} &
0.708&\textbf{0.445}&0.814&0.907&0.740&15.081&6.833&0.708&90.572
\\
\hline
\multirow{2}[4]{*}{RB} & \multicolumn{1}{|c|}{$\widehat\beta$} & 10.346&\textbf{1.646}&34.182&9.509&3.964&-272.773&-29.818&10.346&120.700
\\
\cline{2-11}          & \multicolumn{1}{|c|}{$\widehat\sigma$} &
12.200&\textbf{-2.660}&51.905&10.660&2.955&70.980&76.600&12.200&456.115
\\
\hline
\end{tabular}
}
\end{center}
\end{table}

\end{landscape}
 \clearpage
}

\subsection{Results}

To get a better understanding of the performance of the estimators considered in this study, we present a brief discussion of Tables 1 to 14. We compare the variance, RB, and MSE performance of the estimators to one another in various parameter settings in the Monte Carlo study.

We start by first noting that the RB, variance, and MSE of each of the estimation methods all improve when the sample sizes increase. 
The most notable improvements in overall MSE as the sample size increases from 30 to 500 are in the MLE, MDE.TV and MME, however, as will be mentioned later, the MME method should likely not be considered for estimation even for large samples.
Interestingly, we find that the methods that perform well for smaller sample sizes do not always perform well for larger samples when compared to the competitor methods. In particular, when the sample size is small ($n = 30$ or $n = 50$), one can see that the MDE.SD, MDE.CvM and LME have, in the majority of cases, the smallest variance, RB and MSE values. 
Note that this small sample behaviour of the variance and MSE measures holds for all but the parameter $\beta=1.1$. 
However, when considering the case where the sample size is small and the parameter $\beta$ is specified to be small, there are instances where the MDE.CvM exhibits comparatively larger bias than MDE.SD, LME, and even MLE.b.
It is thus clear that the underlying parameter values also play a role in the relative performance of the various estimation techniques considered. 
Finally, when considering the RB for the various estimators, we note that MDE.SD  most frequently gives rise to negative RB values. Based on the generally positive RB values associated with the remaining estimators, it therefore seems that these other estimators tend to overestimate the true parameter value. 
This behaviour of MDE.SD (and, indeed, all other MDE-type estimators considered) is difficult to explain as their distributional properties depend on the parameters and do not permit simple closed-form expressions.

Note that while
the MDE.KL, MDE.SD and LME are also relatively good estimators (in an MSE sense) when the sample size is large ($n=200$ or $n=500$), the MLE method now has the \emph{best} performance among all methods considered in this setting.







Next, we consider the behaviour of the estimators as the parameter settings of the Monte Carlo change from one value to the next. As $\beta$ increases from 1.1 to 2, the values of the approximated variance, RB, and MSE of the estimators for both $\beta$ and $\sigma$ have the tendency to increase as well. However, these values then immediately start to decrease again when $\beta$ is strictly larger than 2.

When the values of $\sigma$ increases from 0.5 to 2, the simulation approximation of the variance of the estimators of $\sigma$ generally increases for all of the different estimation methods. The same is not necessarily true for the estimation of $\beta$, where we find that there are instances where the approximated variance of the estimator of $\beta$ (using a given estimation method) decreases when the true $\sigma$ increases from 1 to 2, and increases in others. 

When comparing the class of estimators based on minimum distance measures to the remaining methods, one can readily see that the MDE.SD and MDE.CvM estimators are the best performers in terms of the MSE for small to moderate samples. The LME is an excellent competitor, but the MLE and MLE.b estimators perform relatively poorly in terms of the MSE in these cases. However, for larger sample sizes, the MLE and MLE.b clearly outperform the entire MDE class of estimators. Interestingly, the LME remains competitive here too, but is almost never found to have the best MSE performance.

The performances of the estimators within the class of MDE methods also differ wildly from one another. We have already noted that the MDE.SD and MDE.CvM methods generally have good MSE performance for small to moderate samples sizes, but within this class we find that the MDE.TV and MDE.$\chi^2$ have some of the worst performances. The MDE.KL method is something of a mixed bag: some parameter settings yield good MSE performances, where others yield extremely poor MSE performance. Generally, however, we can state that the MDE.KL method is a better performer than MDE.TV and MDE.$\chi^2$, although this still does not place this estimator near the top performers among the estimation methods considered.

The overall worst performing estimator is MME regardless of which value of $\beta$ or $\sigma$ is considered. While this estimator improves with an increase in sample size, it still has the worst performance  when compared to all other estimation methods.
In Figures \ref{fig:accplots1} and \ref{fig:accplots2} the logarithm of the 
TMSE 
measure is plotted against the samples sizes in order to produce a more intuitive representation of the behaviour of a selection of estimators over varying sample sizes. The estimators MLE, MLE.b, LME, MDE.CvM and MDE.SD are selected for includion in these figures since, according to the tables, these estimators generally perform well (the 
TMSE
values of the remaining estimators can be found in the tables). We see that, as expected, the 
TMSE
measure's value generally decreases as sample size increases. We note, however, that the MLE and MLE.b Accuracies are erratic for small samples; 
see, for example, the graphs associated with $\beta=2$ in Figures \ref{fig:accplots1} and \ref{fig:accplots2}.
Furthermore, we see that while the MLE and MLE.b techniques have the worst 
TMSE
performance for small sample sizes, this trend reverses when the samples sizes increase.
Conversely, it also clear that the MDE.CvM and MDE.SD estimators 
performs well, in terms of
TMSE,
for smaller sample sizes, in many cases outperformed only when the sample sizes increase to $n=100$ and higher. \textcolor{red}{It should be noted that the $LME$ also performs well for smaller sample sizes. Furthermore, the numerical results indicate that the relative performance of the estimators depend, to some degree, on the specific parameter values considered.}


\paragraph{Remark:} \it The extremely large values for the variance reported for the MLE and MLE.b estimators can be attributed to the difficulty in estimating the parameters for small sample sizes. The large variances appear since, in a small number of aberrant simulated data sets, the estimation of the parameters produces extremely large estimates for the parameters. However, if we calculate the `trimmed' variance, that is, the variance of the smallest $(100-\alpha)\%$ of the parameter estimates, the variance drastically reduces to more reasonable values. For example, in the case where $n=30$, $\beta=1.5$ and $\sigma=1$, we find that the approximate variance of the MLEs is 537.8, but, trimming only $1\%$ of the highest estimates, the trimmed variance is approximately 5.5. \rm

\section{Practical application}

The various Lomax parameter estimation techniques are now applied to a set of observed losses resulting from wind related catastrophes. This data set is also analysed in \cite{BRAZ2003} and \cite{ALLI2022}. The data set is comprised of the monetary expenses incurred as a result of wind related catastrophes in 40 separate instances during 1977 (these values are rounded to the nearest million US dollars). This rounding causes unrealistic clustering in the data which may lead to problems when fitting the Lomax distribution. In order to circumvent the associated problems, we use the de-grouping algorithm discussed in the two papers mentioned above. This algorithm replaces the values in each group of tied observations with the expected value of the order statistics of the uniform distribution with the same range. That is, if there are $k$ observations in an interval $(l,u)$, we replace the observations by
\begin{equation*}
    \left(\frac{m+1-j}{m+1}\right)l+\left(\frac{j}{m+1}\right)u,
\end{equation*}
for $j \in \{1,\dots,k\}$. We emphasise that this de-grouping algorithm d]oes not change the mean of the data. The de-grouped data can be found in Table \ref{wind}.


\begin{table}[!htbp!]%
\caption{Wind catastrophes data set.} \label{wind}
\begin{center}
\small
\begin{tabular}{rrrrrrrr}
\hline
1.58 & 1.65 & 1.73 & 1.81 & 1.88 & 1.96 & 2.04 & 2.12 \\
2.19 & 2.27 & 2.35 & 2.42 & 2.70 & 2.90 & 3.10 & 3.30 \\
3.75 & 4.00 & 4.25 & 4.70 & 4.90 & 5.10 & 5.30 & 5.70 \\
5.90 & 6.10 & 6.30 & 7.83 & 8.17 & 9.00 & 15.00 & 17.00 \\
22.00 & 23.00 & 23.83 & 24.17 & 25.00 & 27.00 & 32.00 & 43.00 \\
\hline
\end{tabular}
\end{center}
\end{table}

Since the observed (rounded) minimum amount of damage is 2, we conclude that no value less than 1.5 was observed. As a result, we subtract 1.5 from each of the observed values so as to ensure that the support of the distribution is appropriate. Assuming a Lomax distribution for the shifted data, we use each of the estimation techniques to obtain the parameter estimates of the distribution. The resulting parameter estimates are shown in Table \ref{parmests}. The table indicates that, with the exception of MDE.TV and MME, the parameter estimates obtained are closely related to each other. 
{We note that the difference between the estimates arrived at using these estimators on the one hand and the remaining estimators on the other hand is unsurprising given the irregular performance of these two estimators observed in the Monte Carlo study above.}

The code for this example can be found here:
\begin{center}
\href{https://tinyurl.com/LomaxPracticalCode}{\texttt{https://tinyurl.com/LomaxPracticalCode}}
\end{center}

\begin{table}[!htbp!]%
\caption{Estimated parameters for the wind catastrophes data set.} \label{parmests}
\begin{center}
\small
\begin{tabular}{lrr}
Estimation technique & $\widehat{\beta}$ & $\widehat{\sigma}$ \\
\hline
MLE & 6.726 & 53.099 \\
MLE.b & 6.726 & 53.099 \\
LME & 6.367 & 49.512 \\
MDE.CvM & 5.544 & 39.991 \\
MDE.SD & 6.773 & 49.461 \\
MDE.$\chi^2$ & 5.724 & 60.334 \\
MDE.TV & 13.602 & 143.149 \\
MDE.KL & 6.727 & 53.115 \\
MME & 11.944 & 100.958 \\
PWM & 6.367 & 49.512 \\
\hline
\end{tabular}
\end{center}
\end{table}

\begin{figure}[!htbp!]
    \centering
    \includegraphics[width=0.95\textwidth]{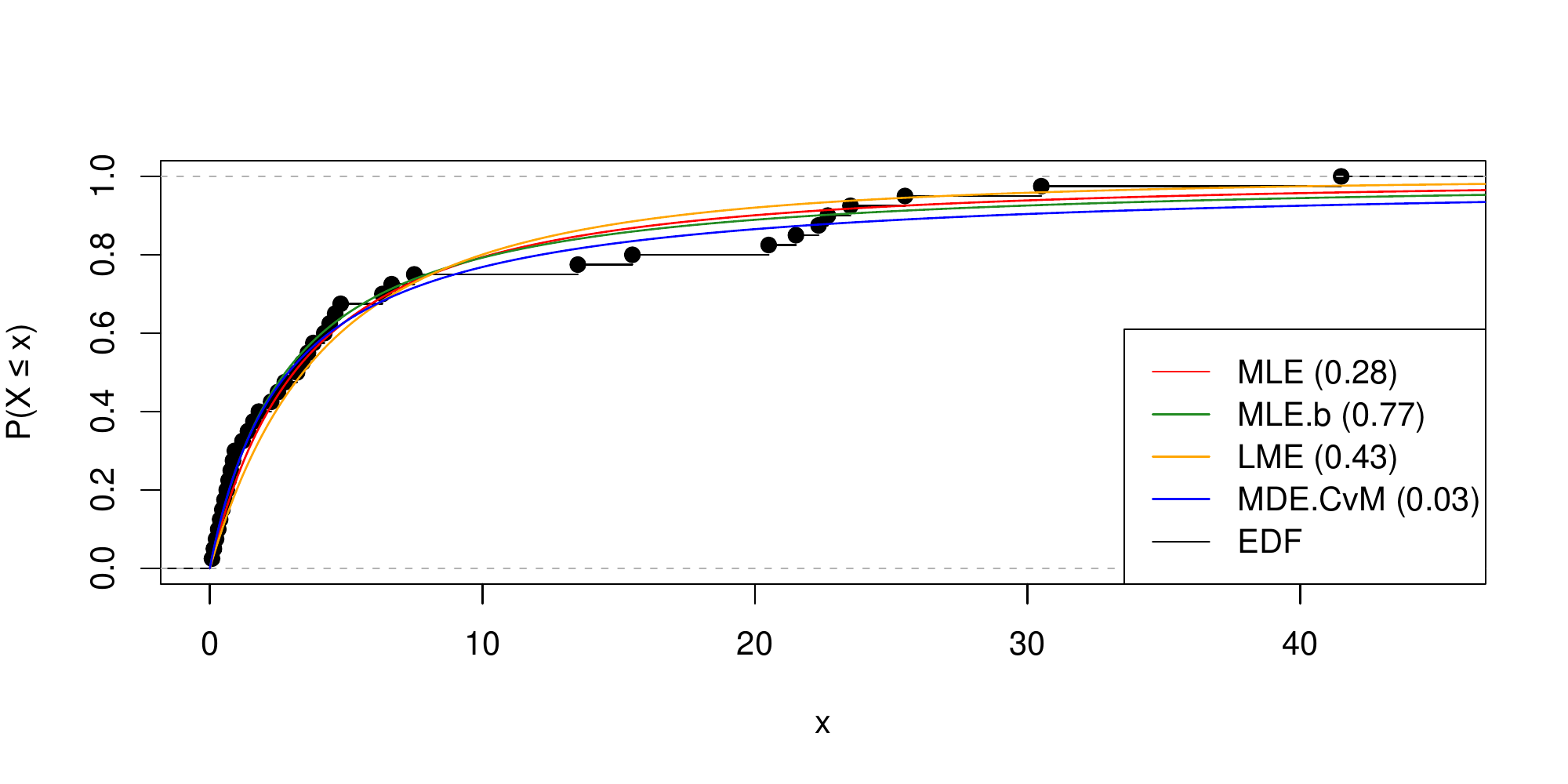}
    \caption{The empirical distribution function and fitted distribution functions based on the wind catastropes data}
    \label{Fig1}
\end{figure}

{Figure \ref{Fig1} shows the EDF of the data with several fitted distribution function obtained using the MLE, MLE.b, LME and MDE.CvM. Visual inspection of Figure \ref{Fig1} indicates that, while minor deviations from the Lomax distribution are present, the Lomax provides a suitable model for the wind catastrophe data.}

{We also perform formal goodness-of-fit tests in order to test the hypothesis that the data is realised from the Lomax distribution. For this purpose, we calculate the $p$-values associated with the Kolmogorov-Smirnov test for each of the fitted distributions. For the sake of brevity, we mention only that we use the parametric bootstrap approach detailed in \cite{ABE2019}. The $p$-values associated with the fitted distributions are 0.28, 0.77, 0.43 and 0.03, respectively. As a result, we are able to use the fitted Lomax model to calculate theoretical properties such as quantiles and moments parametrically.}

\paragraph{Remark:} 
As suggested by a reviewer, due to the instability observed in some of the estimators for smaller sample sizes, we include a small Monte Carlo simulation similar to those displayed in Section 3. We specify the sample size to be $n=40$ and the parameter values to be $\beta=6$ and $\sigma=50$, as these values closely correspond to the majority of the estimated values obtained for the wind catastrophe data set. The results from the study are displayed in Table \ref{tbl:n40sig50}. When analysing the wind catastrophe data we are able to obtain estimators for the parameters using the MLE and MLE.b methods. In contrast, in the simulation study, we find that the optimisers required for these estimators very rarely converge, meaning that these estimates often do not exist. We thus exclude the MLE and MLE.b from the table and report only the remaining estimators. We must therefore caution the user against these methods in the case of small samples. Similar erratic tendencies for these estimators were observed in Figures \ref{fig:accplots1} and \ref{fig:accplots2}.

      \begin{table}[htbp]
  \centering
  \caption{Comparison of different estimation methods for $\beta=6,  \sigma=50$, and  $n=40$.} 
  \label{tbl:n40sig50}
  \resizebox{\columnwidth}{!}{
    \begin{tabular}{|cc|r|r|r|r|r|r|r|}
\cline{2-9}    \multicolumn{1}{r|}{} & \textbf{METHODS} & \multicolumn{1}{l|}{\textbf{LME}} & \multicolumn{1}{l|}{\textbf{MDE.CvM}} & \multicolumn{1}{l|}{\textbf{MDE.SD}} & \multicolumn{1}{l|}{\textbf{MDE.$\chi^2$}} & \multicolumn{1}{l|}{\textbf{MDE.TV}} &\multicolumn{1}{l|}{\textbf{MDE.KL}} & \multicolumn{1}{l|}{\textbf{MME}}  \\
    \hline
\multirow{2}[4]{*}{Mean} & \multicolumn{1}{|c|}{$\widehat\beta$} &
6.736&8.887&6.043&9.188&30.006&11.417&33.736
\\
\cline{2-9}          & \multicolumn{1}{|c|}{$\widehat\sigma$} &
54.338&73.759&49.179&259.491&428.293&97.937&320.381
\\
\hline
\multirow{2}[4]{*}{Var} & \multicolumn{1}{|c|}{$\widehat\beta$} & 
56.331&90.114&60.306&2350.520&1514.388&304.023&426080.072
\\
\cline{2-9}          & \multicolumn{1}{|c|}{$\widehat\sigma$} &
4259.071&6856.270&4582.387&212172.392&330753.014&24748.052&52987947.302
\\
\hline
\multirow{2}[4]{*}{RB} & \multicolumn{1}{|c|}{$\widehat\beta$} & 
12.273 & 48.118 & 0.712 & 53.137 & 400.092 & 90.288 & 462.263
\\
\cline{2-9}          & \multicolumn{1}{|c|}{$\widehat\sigma$} &
8.677 & 47.519 & -1.642 & 418.983 & 756.585 & 95.873 & 540.762
\\
\hline
\multirow{2}[4]{*}{MSE} & \multicolumn{1}{|c|}{$\widehat\beta$} & 
56.868&98.440&60.302&2360.450&2090.500&333.340&426806.740
\\
\cline{2-9}          & \multicolumn{1}{|c|}{$\widehat\sigma$} &
4277.465&742.089&4582.603&256037.782&473825.260&27043.486&53055754.526
\\
\hline
\multicolumn{2}{|c|}{TMSE} & 
4334.333&7518.530&4642.905&258398.232&475915.760&27376.827&53482561.267\\
\hline
\end{tabular}
}
\end{table}

\section{Conclusion}




In this study different parameter estimation methods for the two parameters of the Lomax distribution are explored and discussed. We review $L$-moment estimators, probability weighted moments estimators, maximum likelihood estimators (with and without a bias adjustment), method of moments estimators, and three different minimum distance estimators. The specific goals of the study were 
{to provide an overview of the different estimation methods} as well as to determine whether, for this distribution, the minimum distance estimators are feasible estimation alternatives when compared to traditional methods of estimation like maximum likelihood.
In an attempt to ascertain the properties of these estimators, all of the methods were compared to one another 
 via a comprehensive Monte Carlo simulation,
assuming different parametric values for small to large samples. 
Unsurprisingly, this study showed that the MLEs perform well in large samples, but yield severely biased answers in small samples. The bias correction outlined in \cite{GILE2011}, however, noticeably reduces the bias of the MLE estimates for moderate to large sample sizes.
The traditional MME is found to be a universally poor performer in terms of MSE for all settings of the parameters $\beta$ and $\sigma$, and the given sample sizes. The simulation shows that LME is stable under different parameter settings, including varying values of $\beta$, $\sigma$, and the sample size. So, while this estimator is almost never found to have the best  performance in either small or large sample settings, it can be considered to be a `safe' option for the estimation of the parameters of the Lomax distribution.

Finally, several estimation methods based on distance measures were investigated and were found to produce varied results. These methods all attempt to estimate the parameters of the Lomax distribution by considering different measures of distance measures that need to be minimised. The measures considered include Cram\'{e}r-von Mises (CVM), squared difference, and some phi-divergence measures (including the Kullback-Leibler (KL) divergence, total variation (TV) distance and chi-square divergence). The results of the simulation study established that for this class of MDE methods, the MDE.SD and MDE.CvM methods have the best overall MSE performance for small to moderate samples sizes, and we would recommend either of these methods for the estimation of the Lomax parameters $\sigma$ and $\beta$ when sample sizes are smaller than 100. However,  within this MDE class we also find two of the worst performing methods, namely the MDE.TV and MDE.$\chi^2$ methods.  The MDE.KL method produces reasonably good MSE performance for the moderate to large sample sizes, but, as with MLE, the quality of the estimator fluctuates tremendously for smaller samples. 

The results show that no \emph{single} estimator has the best overall performance, but we can identify the distance-based measure, MDE.SD, as generally having the best MSE performance for small sample sizes ($n = 30$ and $n = 50$), whereas  MLE.b is found to be the better choice for moderate and large sample sizes ($n = 100$, $n = 200$ and $n = 500$), for almost all $\sigma$ and $\beta$ values considered.


In the literature one finds competing views about the usefulness of MDEs and MLEs. In the case of multivariate data, \cite{LEIB2022} (specifically estimators for the parameters of a copula) advocates for the use of MDEs because MLEs have the disadvantage that one needs the density of the copula, whereas for MDEs one only requires the empirical distribution estimate of the copula. However, in an extensive overview paper comparing MDEs and MLEs for parameter estimation of copulas, \cite{WEIS2011} found that MLEs, in addition to having a lower computational cost, also produce lower estimation biases. Naturally, the other main advantages of the traditional MLEs are the ease with which asymptotic standard errors can be obtained, the fact that they have asymptotically normal distributions, and, consequently, confidence intervals for the parameter being estimated are easily obtained. In contrast, the standard errors of the other estimators considered in this paper require either complex analytical derivations (possibly with no tractable forms) or require the use of resampling methods.


\bibliography{ThobekaBIB.PhD}

\appendix
\section*{Appendix: Additional tables}

Tables 12 to 17 report the results of the Monte Carlo experiments for sample sizes $n=200$ and $n=500$. Each table shows the empirical expected value, variance, RB, MSE and TMSE for the various estimators considered, as before.

\afterpage{%
    \clearpage%
\begin{landscape}
      \begin{table}[htbp]
  \centering
\caption{Comparison of different estimation methods for different values of $\beta$ ($\sigma=0.5$ and $n=200$). \label{tbl:n200sig05}}

}
\end{table}
     \begin{table}
   \caption{Comparison of different estimation methods for different values of $\beta$ ($\sigma=1$ and $n=200$). \label{tbl:n200sig1}}
  \begin{center}
  \resizebox{1.3\textwidth}{!}{
    \begin{tabular}{|cc|r|r|r|r|r|r|r|r|r|}
\cline{2-11}    \multicolumn{1}{r|}{} & \textbf{METHODS} & \multicolumn{1}{l|}{\textbf{MLE}} & \multicolumn{1}{l|}{\textbf{MLE.b}} & \multicolumn{1}{l|}{\textbf{LME}} & \multicolumn{1}{l|}{\textbf{MDE.CvM}} & \multicolumn{1}{l|}{\textbf{MDE.SD}} & \multicolumn{1}{l|}{\textbf{MDE.$\chi^2$}} & \multicolumn{1}{l|}{\textbf{MDE.TV}} &\multicolumn{1}{l|}{\textbf{MDE.KL}} & \multicolumn{1}{l|}{\textbf{MME}} \\
\cline{2-11}
    \multicolumn{11}{c}{$\beta=1.1,  \sigma=1$ $n=200$}\\
    \hline
\multirow{2}[4]{*}{Mean} & \multicolumn{1}{|c|}{$\widehat\beta$} & 1.148&1.107&1.366&1.159&1.130&0.419&0.650&1.148&2.235
\\
\cline{2-11}          & \multicolumn{1}{|c|}{$\widehat\sigma$} &
1.074&1.008&1.432&1.088&1.050&1.291&1.411&1.074&5.635
\\
\hline
\multirow{2}[4]{*}{Var} & \multicolumn{1}{|c|}{$\widehat\beta$} & 0.034&\textbf{0.029}&0.036&0.067&0.060&14.728&0.049&0.034&0.042
\\
\cline{2-11}          & \multicolumn{1}{|c|}{$\widehat\sigma$} &
0.083&\textbf{0.069}&0.109&0.137&0.124&1.689&0.216&0.083&5.172
\\
\hline
\multirow{2}[4]{*}{RB} & \multicolumn{1}{|c|}{$\widehat\beta$} & 4.373&\textbf{0.636}&24.218&5.355&2.746&-61.927&-40.946&4.373&103.209
\\
\cline{2-11}          & \multicolumn{1}{|c|}{$\widehat\sigma$} &
7.440&\textbf{0.830}&43.220&8.770&5.020&29.130&41.060&7.440&463.530
\\
\hline
\end{tabular}
}
\end{center}
\end{table}

   \begin{table}
   \caption{Comparison of different estimation methods for different values of $\beta$ ($\sigma=2$ and $n=200$). \label{tbl:6}}
  \begin{center}
  \resizebox{1.3\textwidth}{!}{
    \begin{tabular}{|cc|r|r|r|r|r|r|r|r|r|}
\cline{2-11}    \multicolumn{1}{r|}{} & \textbf{METHODS} & \multicolumn{1}{l|}{\textbf{MLE}} & \multicolumn{1}{l|}{\textbf{MLE.b}} & \multicolumn{1}{l|}{\textbf{LME}} & \multicolumn{1}{l|}{\textbf{MDE.CvM}} & \multicolumn{1}{l|}{\textbf{MDE.SD}} & \multicolumn{1}{l|}{\textbf{MDE.$\chi^2$}} & \multicolumn{1}{l|}{\textbf{MDE.TV}} &\multicolumn{1}{l|}{\textbf{MDE.KL}} & \multicolumn{1}{l|}{\textbf{MME}} \\
\cline{2-11}
      \multicolumn{11}{c}{$\beta=1.1,  \sigma=2$ $n=200$}\\
    \hline
\multirow{2}[4]{*}{Mean} & \multicolumn{1}{|c|}{$\widehat\beta$} & 1.155&1.113&1.380&1.156&1.127&-0.656&0.666&1.155&2.250
\\
 \cline{2-11}          & \multicolumn{1}{|c|}{$\widehat\sigma$} &
2.128&1.996&2.850&2.128&2.056&2.856&2.806&2.128&11.271
\\
\hline
\multirow{2}[4]{*}{Var} & \multicolumn{1}{|c|}{$\widehat\beta$} & 0.032&\textbf{0.027}&0.036&0.056&0.051&134.211&0.049&0.032&0.042
\\
 \cline{2-11}          & \multicolumn{1}{|c|}{$\widehat\sigma$} &
0.309&\textbf{0.257}&0.425&0.472&0.429&15.687&0.824&0.309&58.604
\\
\hline
\multirow{2}[4]{*}{RB} & \multicolumn{1}{|c|}{$\widehat\beta$} &
 4.964&\textbf{1.200}&25.455&5.055&2.491&-159.627&-39.500&4.964&104.582
\\
\cline{2-11}          & \multicolumn{1}{|c|}{$\widehat\sigma$} &
6.380&\textbf{-0.185}&42.480&6.410&2.790&42.775&40.310&6.380&463.560
\\
\hline
\end{tabular}
 }
 \end{center}
\end{table}

      \begin{table}[htbp]
  \centering
\caption{Comparison of different estimation methods for different values of $\beta$ ($\sigma=0.5$ and $n=500$). \label{tbl:n500sig05}}
}
\end{table}
      \begin{table}[htbp]
  \centering
\caption{Comparison of different estimation methods for different values of $\beta$ ($\sigma=1$ and $n=500$). \label{tbl:n500sig1}}
}
\end{table}

      \begin{table}[htbp]
  \centering
\caption{Comparison of different estimation methods for different values of $\beta$ ($\sigma=2$ and $n=500$). \label{tbl:n500sig2}}
}
\end{table}
\end{landscape}
 \clearpage
}


\end{document}